\documentclass[secnumarabic,amssymb, nobibnotes, nofootinbib, aps, prd,showpacs]{revtex4}

\usepackage{epsfig}
\usepackage{amssymb}
\usepackage{graphicx}
\usepackage{amsmath}
\renewcommand{\text}[1]{%
\ifthenelse{\equal{#1}{r1}}{r_1}{}%
\ifthenelse{\equal{#1}{r2}}{r_2}{}%
\ifthenelse{\equal{#1}{fB}}{f_B}{}%
\ifthenelse{\equal{#1}{fP}}{f_P}{}%
\ifthenelse{\equal{#1}{mB}}{m_B}{}%
\ifthenelse{\equal{#1}{xi0}}{\sigma_0}{}%
\ifthenelse{\equal{#1}{s0}}{s_0}{}%
\ifthenelse{\equal{#1}{M2}}{M^2}{}%
\ifthenelse{\equal{#1}{barxi}}{\bar\sigma}{}%
\ifthenelse{\equal{#1}{uu}}{u}{}%
\ifthenelse{\equal{#1}{ubar}}{\bar{\sigma}}{}%
\ifthenelse{\equal{#1}{vv}}{v}{}
\ifthenelse{\equal{#1}{phiBp}}{\phi^B_+}{}%
\ifthenelse{\equal{#1}{PhiBp}}{\phi^B_+}{}%
\ifthenelse{\equal{#1}{phiBmin}}{\phi^B_-}{}%
\ifthenelse{\equal{#1}{PhiBmin}}{\phi^B_-}{}%
\ifthenelse{\equal{#1}{PhiBpm}}{\overline\Phi^B_{\pm}}{}%
\ifthenelse{\equal{#1}{ppsiV}}{\Psi_V^B}{}%
\ifthenelse{\equal{#1}{ppsiA}}{(\Psi_A^B}{}%
\ifthenelse{\equal{#1}{bbarXA}}{\Psi_V^B}{}%
\ifthenelse{\equal{#1}{bbarYA}}{\overline Y_A^B}{}%
\ifthenelse{\equal{#1}{fV}}{f_V}{}%
\ifthenelse{\equal{#1}{mV}}{m_V}{}%
\ifthenelse{\equal{#1}{mP}}{m_P}{}%
\ifthenelse{\equal{#1}{lb}}{\lambda_B}{}%
}

\newcommand{\ba}{\begin{eqnarray}}
\newcommand{\ea}{\end{eqnarray}}
\newcommand{\be}{\begin{equation}}
\newcommand{\ee}{\end{equation}}

\makeatletter
\def\fmslash{\@ifnextchar[{\fmsl@sh}{\fmsl@sh[0mu]}}
\def\fmsl@sh[#1]#2{%
  \mathchoice
    {\@fmsl@sh\displaystyle{#1}{#2}}%
    {\@fmsl@sh\textstyle{#1}{#2}}%
    {\@fmsl@sh\scriptstyle{#1}{#2}}%
    {\@fmsl@sh\scriptscriptstyle{#1}{#2}}}
\def\@fmsl@sh#1#2#3{\m@th\ooalign{$\hfil#1\mkern#2/\hfil$\crcr$#1#3$}}
\makeatother

\begin{document}

\title{$B, B_{s}\to K$ form factors: an update of light-cone sum rule results
}

\author{Goran Duplan\v ci\'c}
\email[]{gorand@thphys.irb.hr}
\author{Bla\v zenka Meli\'c}
\email[]{melic@thphys.irb.hr}
\affiliation{Theoretical Physics Division, Rudjer Boskovic Institute, Bijeni\v cka 54, \\HR-10002 Zagreb, Croatia}
\date{May 29, 2008}%

\begin{abstract}
We present an improved QCD light-cone sum rule (LCSR) calculation of the $B \to K$ and $B_{s} \to K$ form factors 
by including SU(3)-symmetry breaking corrections. 
We use recently updated $K$ meson distribution amplitudes which incorporate the complete SU(3)-breaking structure. 
By applying the method of direct integration in the complex plane, which is presented in detail, 
the analytical extraction of the imaginary parts of LCSR hard-scattering amplitudes becomes unnecessary and therefore the complexity 
of the calculation is greatly reduced. 
The values obtained for the relevant $B_{(s)} \to K$ form factors are as follows:
$f^+_{BK}(0)= 0.36^{+0.05}_{-0.04}$, $f^+_{B_sK}(0)= 0.30^{+0.04}_{-0.03}$, and $f^T_{BK}(0)= 0.38\pm 0.05$, $f^T_{B_sK}(0)= 0.30\pm 0.05$. 
By comparing with the 
$B\to \pi$ form factors extracted recently by the same method, we find the following SU(3) violation among the 
$B \to $ light form factors: 
$f^+_{BK}(0)/f^+_{B\pi}(0) = 1.38^{+0.11}_{-0.10}$, $f^+_{B_sK}(0)/f^+_{B\pi}(0) = 1.15^{+0.17}_{-0.09}$, 
$f^T_{BK}(0)/f^T_{B\pi}(0) = 1.49^{+0.18}_{-0.06}$, and 
$f^T_{B_sK}(0)/f^T_{B\pi}(0) = 1.17^{+0.15}_{-0.11}$. 
\end{abstract}

\pacs{13.25.Hw,12.38.Lg}
\keywords{B-decays, QCD, Sum rules}
\maketitle

\section{Introduction} 

The $B \to$ light meson form factors are important ingredients in the analysis of semileptonic $B$ decays, as well as of nonleptonic 
two-body $B$ decays, where they serve for extraction of the Cabibbo-Kobayashi-Maskawa matrix elements. 
They have been studied by light-cone sum rule (LCSR) \cite{lcsr} methods in several papers \cite{BKR,BBKR,KRWY,BBB,Ball98,KRW,KRWWY,
Huang,BZ01,BZ04,KMO,DFFH}, and most recently in \cite{DKMMO}. 
In this paper we want to concentrate on the flavor SU(3)-symmetry breaking corrections in $B \to K$ and $B_s \to K$ form factors, closely 
following the method presented in \cite{DKMMO}. In \cite{DKMMO}, the $B \to \pi $ form factors were analyzed, 
and in contrast to the previous calculations 
with the pole mass for $m_b$, the $\overline{MS}$ mass $\overline{m}_b(\mu)$ was used. This choice more naturally follows the idea of 
the perturbative calculation of the hard-scattering amplitudes. Since the sum rule calculation of $f_B$ and $f_{B_s}$ decay constants is also 
available in the $\overline{MS}$ scheme \cite{JL}, we are able to consistently perform estimation of the $f_{B_{(s)}K}$ form factors in 
this scheme. 

The notion of SU(3) breaking is particularly interesting in a view of discrepancies of measured values for 
$B_{(s)} \to \pi K$ decay widths and CP asymmetries compared to Standard Model predictions. The $B_{(s)} \to K$ transition form factors enter different models 
for calculating these decays, and according to 
the recent analysis \cite{CGR}, one solution of these discrepancies is given by assuming the large SU(3)-breaking effects, either 
in strong phases or in amplitudes. Our intention is to calculate these effects in different $B_{(s)} \to K$ form factors by 
using all known SU(3)-breaking corrections in the parameters and in distribution amplitudes (DAs) entering the LCSR calculation. 

%
Up to now, in \cite{Huang,KMM,WHF}, the main SU(3)-breaking 
effects were included by considering SU(3)-breaking in the parameters of the leading twist DAs, such as $f_K/f_{\pi}$ and $\mu_K/\mu_{\pi}$, 
and by inserting $p^2 = m_K^2 \neq 0$ at LO. 
In the meantime, the complete SU(3)-symmetry breaking corrections in the $K$ meson DAs are known \cite{BBL}. In 
\cite{BBL}, the 
authors complete the analysis of SU(3)-breaking corrections done in \cite{BF,BallDA,BK,BL,BGG}, for all twist-3 and twist-4 two- and 
three-particle DAs, by including also $G$-parity-breaking corrections in $m_s - m_q$. 
Therefore our analysis will include complete SU(3)-breaking effects in both kaon DAs, as well as in the hard-scattering amplitudes at LO. 
At next-to-leading order (NLO) in the hard-scattering amplitudes, the inclusion of
 $m_s$ and $m_K^2$ effects complicates the calculation. Because of the complexity of mixing between twist-2 and twist-3 DAs, 
we were not able to perform consistent calculations with $m_s$ included in the quark propagators. 
Therefore, in those amplitudes we also set $p^2=m_K^2 =0$, 
and consistently use twist-2 and twist-3 two-particle kaon DAs without mass corrections. However, we analyze the kaon mass effects ($p^2 =m_K^2$) at 
NLO and include them in the error estimates. More detailed discussion about this point will be given in Sec.2. 

Since the LO hard-scattering amplitudes are already complicated when the twist-4 and three-particle DAs are included, we will use the new, 
numerical method to calculate the sum rules. The idea is to use the analyticity of the integrals, and to continue 
them to the complex plane. The integrals are then performed over the contour in a complex plane, 
and the imaginary part is obtained numerically. 
The details of the method will be given below.

\section{LCSR for $B_{(s)} \to K$ form factors} 

To obtain the form  factors $f^+_{BK}$, $f^0_{BK}$, and $f^T_{BK}$ from LCSR we consider the 
vacuum-to-kaon correlation function of a weak current and a current with the $B$ meson quantum numbers:
\ba
F_{\mu}(p,q)&=&i\int d^4x ~e^{i q\cdot x}
\langle K(p)|T\left\{\bar{s}(x)\Gamma_\mu b(x), 
m_b\bar{b}(0)i\gamma_5 d(0)
\right\}|0\rangle
\nonumber\\
&=&\Bigg\{\begin{array}{ll}
F(q^2,(p+q)^2)p_\mu +\widetilde{F}(q^2,(p+q)^2)q_\mu\,,& ~~\Gamma_\mu= \gamma_\mu\\
&\\
F^T(q^2,(p+q)^2)\big[p_\mu q^2-q_\mu (q p)\big]\,,& ~~\Gamma_\mu= -i\sigma_{\mu\nu}q^\nu\\
\end{array} 
\label{eq:corr}
\ea
for  the two different $b\to s$ transition currents.  
For definiteness, we consider the 
$\bar{B}_d\to \overline{K}^0(s\overline{d})$ flavor configuration, and use the isospin symmetry limit, ignoring replacement of 
a $u$ quark by a $d$ quark in the penguin current. For the case of $f^{+,0,T}_{B_{s}K}$ form factors we consider 
the $B_s\to K^0(\overline{s}d)$ decay. This enables us to use the same correlation function, 
with $s$ and $d$ quarks interchanged, but in the kaon DAs 
one has to take care about the fact that DAs from \cite{BBL} are defined for the configuration in which the momentum fraction 
carried by the $s$-quark is $u$ (i.e., $\alpha_1$ in the three-particle DAs), and $\bar{u} = 1-u$ ($\alpha_2$ in the three-particle DAs) is the 
antiquark momentum fraction. 
Since we want to explore the SU(3)-breaking corrections we will keep the kaon mass ($p^2 = m_K^2$) and the $m_s$ quark mass in the DAs. 
The light quark masses will be systematically neglected, except in the ratio $\mu_K = m_K^2/(m_s + m_d)$.   

For the large virtualities of the currents above, the correlation function is dominated by the distances $x^2 =0$ near the light cone, and 
factorizes to the convolution of the nonperturbative, universal  part (the light-cone DA) and the perturbative, short-distance part, 
the hard-scattering amplitude, as a sum of contributions of increasing twist.
In contrast to the pion DA, where due to the G-parity odd Gegenbauer moments vanish,
the lowest twist-2 DA of a kaon has an expansion 
\be
\phi_K(u,\mu) = 6 u (1-u) \left ( 1 + a_1^K(\mu) C_1^{3/2} (2 u -1) + a_2^K(\mu) C_2^{3/2}(2 u -1) + ... \right) \, , 
\ee
where we neglect higher moments $a_{>2}^K$. 
\\
We calculate here contributions up to the twist-4 in the 
leading order ($O(\alpha_s^0)$) and up to the twist-3 in NLO, neglecting the three-particle contributions at this level.
Schematically, the contributions are shown in Fig.1 and Fig.2. 
\begin{figure}[t]
\begin{center}
\includegraphics[width=4.5cm]{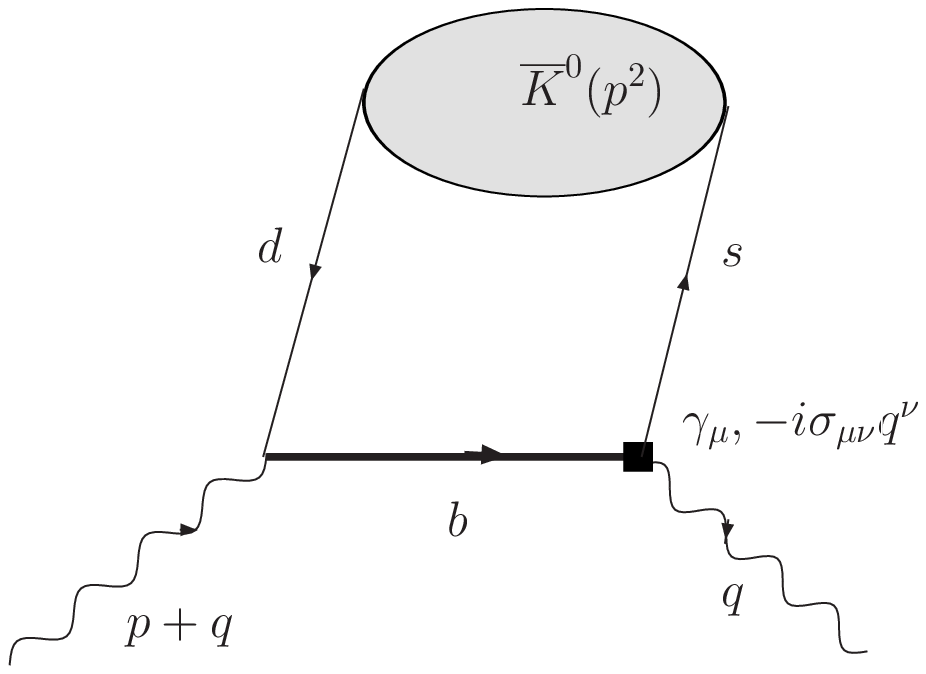}\hspace{1cm}
\includegraphics[width=4.5cm]{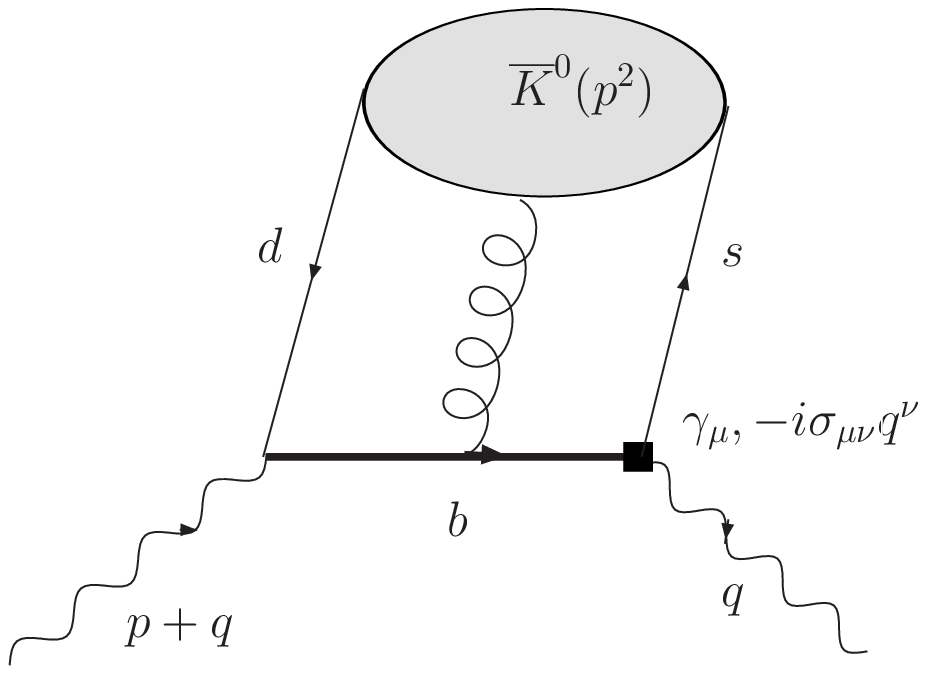}\\
\caption{ \it Diagrams corresponding to the leading-order terms in 
the hard-scattering amplitudes involving the  
two-particle (left) and three-particle (right) kaon DA's shown by ovals. 
Solid, curly, and wavy lines represent quarks, gluons, and external 
currents, respectively. In the case of the $B_s \to K$ transition, $s$ and $d$ quarks are interchanged, and $\bar{K}^0$ is replaced by $K^0$.}
\label{fig-diags}
\end{center}
\end{figure}
\begin{figure}[t]
\begin{center}
\includegraphics[width=11cm]{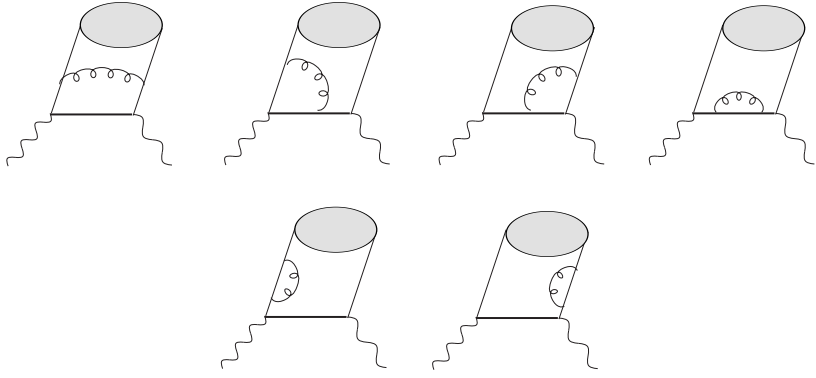}\\
\caption{ \it Diagrams contributing to the hard-scattering amplitudes at $O(\alpha_s)$.}
\label{fig-alphas}
\end{center}
\end{figure}
 
By using the hadronic dispersion relation in the virtuality $(p+q)^2$ of the current in the $B$ channel, we can relate the 
correlation function (\ref{eq:corr})
to the $B \to K$ matrix elements, 
\be
\langle K(p)|\bar{s} \gamma_\mu b |\bar B_d(p+q)\rangle=
2f^+_{BK}(q^2)p_\mu +\left(f^+_{BK}(q^2)+f^-_{BK}(q^2)\right)q_\mu\,,
\label{eq:fplBpi}
\ee
\be
\langle K(p)|\bar{s} \sigma_{\mu \nu}q^\nu b
|\bar B_d(p+q)\rangle=
\Big [q^2(2p_\mu+q_\mu) - (m_B^2-m_K^2) q_\mu\Big ]
\frac{i f_{BK}^T(q^2)}{m_B+m_K}\,.
\label{eq:fsigBpi}
\ee
Inserting hadronic states with the $B$-meson quantum numbers between the currents
in (\ref{eq:corr}), one isolates the $B$-meson ground-state contributions 
for all three invariant amplitudes $F(q^2,(p+q)^2)$, $\widetilde{F}(q^2,(p+q)^2)$, and $F^T(q^2,(p+q)^2)$ and 
using (\ref{eq:fplBpi}) and (\ref{eq:fsigBpi}) obtains
\ba
f^+_{BK}(q^2) &=& \frac{e^{m_B^2/M^2}}{2m_B^2 f_B}
\Bigg[F_0(q^2,M^2,s_0^B)+
\frac{\alpha_s C_F}{4\pi}F_1(q^2,M^2,s_0^B)
\Bigg]\,,
\label{eq:fplusLCSR}
\\
f^+_{BK}(q^2)+f^-_{BK}(q^2) &=& 
\frac{e^{m_B^2/M^2}}{m_B^2 f_B}
\Bigg[\widetilde{F}_0(q^2,M^2,s_0^B)+
\frac{\alpha_s C_F}{4\pi}\widetilde{F}_1(q^2,M^2,s_0^B)
\Bigg]\,,
\label{eq:fplminLCSR}
\\
f^T_{BK}(q^2) &=&
\frac{(m_B+m_K)e^{m_B^2/M^2}}{2m_B^2 f_B}
\Bigg[F^T_{0}(q^2,M^2,s_0^B)+
\frac{\alpha_s C_F}{4\pi} F^T_{1}(q^2,M^2,s_0^B)\Bigg]\,.
\nonumber \\
\label{eq:fTLCSR}
\ea
The scalar $B \to K$ form factor is then a combination of the vector form factor (\ref{eq:fplusLCSR}) and 
the $f^-_{BK}$ form factor from (\ref{eq:fplminLCSR}), 
\be
f^0_{BK}(q^2) = f^+_{BK}(q^2) + \frac{q^2}{m_B^2-m_K^2} f^-_{BK}(q^2) \,.
\label{eq:f0}
\ee
In the above, $F_{0(1)}$ and $\widetilde{F}_{0(1)}$ represent the LO (NLO) contributions and  
$f_B=\langle \bar{B}_d |m_b\bar{b}i\gamma_5 d |0 \rangle/m_B^2$ is the $B$-meson decay constant. As usual, the quark-hadron 
duality is used to approximate the heavier state contribution by introducing the effective threshold parameter $s_0^B$, and 
the ground-state contribution of $B$ meson is enhanced by 
the Borel-transformation in the variable $(p+q)^2 \to M^2$. 
Completely analogous relations are valid for $B_s \to K$ form factors, with the replacement $s \leftrightarrow d$ 
in (\ref{eq:fplBpi}) and (\ref{eq:fsigBpi}) and by replacing $m_B$ by $m_{B_s}$, $f_B$ by $f_{B_s}$, as well as $M^2$ by $M_s^2$ and 
$s_0^B$ by $s_0^{B_s}$ in (\ref{eq:fplusLCSR} - \ref{eq:fTLCSR}). 
In addition, in the derivation of the above expressions for $B_s$, one has to take into account that 
$\langle B_s | \bar{b}i\gamma_5 s |0 \rangle/m_{B_s}^2 = f_{B_s}/(m_b + m_s)$.  

The calculation will be performed in the $\overline{MS}$ scheme.  
The $B$ and $B_s$ decay constants $f_{B_{(s)}}$ will be calculated in the $\overline{MS}$ scheme using the sum rule expressions 
from  \cite{JL} with $O(\alpha_s, m_s^2)$ accuracy. 

Each form factor can be written in a form of the dispersion relation:
\ba
F(q^2,M_{(s)}^2,s_0^{B_{s}}) = \frac{1}{\pi}\int\limits_{m_b^2}^{s_0^{B_{(s)}}}
ds e^{-s/M^2_{(s)}}\,\mbox{Im}_s F(q^2,s) \, , 
\label{eq:dispE}
\ea
where now $s = (p + q)^2$. 
The leading-order parts of the LCSR for $f^+_{BK}$, $f^+_{BK}+f^-_{BK}$, and $f^T_{BK}$ form factors have the following forms:
\ba
F_0(q^2,(p+q)^2)&=& m_b^2 f_K \int \limits_0^1\frac{du}{m_b^2-(q+up)^2}\Bigg\{
\varphi_K(u)
+\frac{\mu_K}{m_b}u \, \phi^p_{3K}(u)
\nonumber
\\
& & +\frac{\mu_K}{6m_b}
\Bigg[ 2+\frac{m_b^2+q^2 - u^2 p^2}{m_b^2-(q+up)^2}\Bigg]\phi^\sigma_{3K}(u)
-\frac{m_b^2\phi_{4K}(u)}{2\big(m_b^2-(q+up)^2\big)^2}
\nonumber \\
&& -\frac{u}{m_b^2-(q+up)^2}\int\limits_0^u dv \psi_{4K}(v)\Bigg\}
\nonumber
\\
& & +\int\limits _0^1 dv
\int \frac{{\cal D}\alpha}{\big[m_b^2- \big(q+X p\big)^2\big]^2}
\Bigg \{
m_b f_{3K} \left ( 4 v(q\cdot p) - (1-2 v) X p^2 \right )\Phi_{3K}(\alpha_i)
\nonumber
\\
& & +m_b^2 f_K \bigg [
3 ( \Psi_{4K}(\alpha_i) + \widetilde{\Psi}_{4K}(\alpha_i) )
+ \frac{4 v(1-v) (q\cdot p + X p^2)}{m_b^2 - (q + X p)^2} \Xi_{4K}(\alpha_i)
\nonumber
\\
&& - \left ( 1 - \frac{X p^2}{q\cdot p + X p^2}
\right )
(\Psi_{4K}(\alpha_i) +\Phi_{4K}(\alpha_i)
+ \widetilde{\Psi}_{4K}(\alpha_i) + \widetilde{\Phi}_{4K}(\alpha_i))
\bigg ]
\Bigg  \}
\nonumber
\\
&&  - m_b^2 f_K \int_0^1 dv \int  {\cal D}\alpha  \int_0^X d\xi \frac{1}{\big[m_b^2- \big(q+(X-\xi) p\big)^2\big]^2}
\nonumber
\\
&&
\frac{p^2 q\cdot p}{(q\cdot p + (X-\xi) p^2)^2}(\Psi_{4K}(\alpha_i) +\Phi_{4K}(\alpha_i)
+ \widetilde{\Psi}_{4K}(\alpha_i) + \widetilde{\Phi}_{4K}(\alpha_i))
\,,
\label{eq:corrF}
\ea
\ba
\widetilde{F}_0(q^2,(p+q)^2)&=& m_b f_K\int \limits_0^1\frac{du}{m_b^2-(q+up)^2}\Bigg\{
\mu_K\phi^p_{3K}(u) \qquad\qquad
\nonumber
\\
&& +\frac{\mu_K}{6}
\Bigg[ 1-\frac{m_b^2-q^2 + u^2 p^2}{m_b^2-(q+up)^2}\Bigg]
\frac{\phi^\sigma_{3K}(u)}{u}
-\frac{m_b}{m_b^2-(q+up)^2}\int\limits_0^u dv \psi_{4K}(v)\Bigg\}
\nonumber
\\
&&
+ m_b f_{3K}
\int\limits _0^1 dv
\int \frac{{\cal D}\alpha}{\big[m_b^2- \big(q+X p\big)^2\big]^2}
 ( 2 v-3) p^2 \Phi_{3K}(\alpha_i)
\nonumber
\\
&&  +  4 m_b^2 f_K \int_0^1 dv \int {\cal D}\alpha \int_0^{X} d\xi \frac{1}{\big[m_b^2- \big(q+(X-\xi) p\big)^2\big]^3}
\nonumber
\\
&&
\qquad p^2 (\Psi_{4K}(\alpha_i) +\Phi_{4K}(\alpha_i)
+ \widetilde{\Psi}_{4K}(\alpha_i) + \widetilde{\Phi}_{4K}(\alpha_i)) \, , 
\label{eq:corrFtilde}
\ea
\ba
F^T_0(q^2,(p+q)^2)&=& m_b f_K\int \limits_0^1\frac{du}{m_b^2-(q+up)^2}
\Bigg\{
\varphi_K(u)
+\frac{m_b\mu_K}{3(m_b^2-(q+up)^2)}\phi^\sigma_{3K}(u)
\nonumber
\\
&& -\frac1{2(m_b^2-(q+up)^2)}\Bigg(\frac12 +
\frac{m_b^2}{m_b^2-(q+up)^2}\Bigg)\phi_{4K}(u)
\Bigg\}
\nonumber
\\
&& + m_b f_K \int\limits _0^1 dv
\int \frac{{\cal D}\alpha}{\big[m_b^2- \big(q+X p\big)^2\big]^2}
\nonumber
\\
&& + \bigg \{
2 \Psi_{4K}(\alpha_i) - (1- 2 v) \Phi_{4K}(\alpha_i) +  2 (1 - 2 v) \widetilde{\Psi}_{4K}(\alpha_i) - 
\widetilde{\Phi}_{4K}(\alpha_i)
\nonumber \\
&& \qquad + \frac{4 v(1-v) (q\cdot p + X p^2)}{m_b^2 - (q + X p)^2} \Xi_{4K}(\alpha_i)
\bigg \}
\,,
\label{eq:corrFtens}
\ea
respectively, with $X =\alpha_1 + v \alpha_3$,  ${\cal D}\alpha=
d\alpha_1 d\alpha_2 d\alpha_3 \delta(1-\alpha_1-\alpha_2-\alpha_3)$,
and with the definitions of the twist-2 ($\varphi_K$), twist-3
($\phi^p_{3K}$, $\phi^\sigma_{3K}$, $\Phi_{3K}$),
and twist-4 ($\phi_{4K}$, $\psi_{4K}$,
$\Phi_{4K}$, $\Psi_{4K}$, $\widetilde{\Phi}_{4K}$,
$\widetilde{\Psi}_{4K}$) kaon DA's from \cite{BBL}. 
It is easy to see that for $p^2=0$ the above expressions resemble those given in \cite{DKMMO} for the $B \to \pi$ form factors. 
Here we have also included a contribution from an additional G-parity breaking twist-4 three-particle DA, $\Xi_{4K}$, 
which was first introduced in \cite{BGG}. Its parameter, as well as the rest of DA parameters, are taken from \cite{BBL} where the 
renormalon model is used for describing SU(3)-symmetry breaking for twist-4 DAs. For all details about the 
SU(3)-symmetry breaking effects in the kaon DAs the reader is advised to see \cite{BBL}. 
  
The massless ($m_K^2, m_s \to 0 $) NLO contributions to the LCSR expressions for $B \to K$ form factors, $F_1(p, (p+q)^2)$, etc., 
are the same as those given in the appendix of 
\cite{DKMMO} for $B \to \pi$ form factors.  
All features of these $O(\alpha_s)$ corrections are already listed in \cite{DKMMO}, and we will not repeat them in this paper. 
Unfortunately, we were not able to perform the full NLO calculation with the mass effects included. The problem appeared 
by inclusion of the chirally noninvariant piece of the s-quark propagator, being proportional to $m_s$, in the calculation 
of diagrams from Fig.2. Since now $p^2 = m_K^2$, 
the IR divergences were not present, but there appeared additional UV divergences proportional to $m_s$, 
which have clearly exhibited the mixing among different twists. We could not achieve the cancellation of such 
singularities, since obviously some additional ingredient of mass mixing among twist-2 and twist-3 contributions was missing. 
Although interesting {\it per se}, these mixing effects are nontrivial, and there are beyond a scope of this paper. 
Hence, the repercussions of the mass effects at NLO could only be analyzed by setting $m_s \to 0$. 
We are aware that keeping $O(m_K^2)$ effects, and neglecting the same order effect of $m_s$-proportional terms is not completely 
justified; therefore, we have used the result with $p^2 = m_K^2$ corrections only as an estimation for the neglected mass effects at NLO. 

The final LCSR expressions for $B_{(s)} \to K$ form factors, with the $p^2=m_K^2$ corrections included, have a similar form 
as those for the $p^2=0$ shown in \cite{DKMMO}, but with a more complicated structure now. 
Therefore we are not going to present them here
\footnote{Interested readers can obtain all expressions from the authors in {\it Mathematica} 
\cite{math} form.}. 

\section{Direct integration of the LCSR expressions}

The sum rule expression for the form factors (\ref{eq:dispE}) requires, by definition, calculation of the 
imaginary part of hard-scattering amplitudes. 
Complexity of the extraction of imaginary parts of sum rule amplitudes arises already at the LO level, as one can notice from 
the expressions in Appendix A. One has to be particularly careful about the appearance of the surface terms there. 
At the NLO the results are far more complicated, as one can see in Appendix B of \cite{DKMMO}. The inclusion 
of $p^2=m_K^2$ effects at NLO makes the calculation even more involved. 

Therefore, we would like to present here a method which completely avoids  the use of
explicit imaginary parts of hard-scattering amplitudes,
allowing one to  numerically calculate amplitudes of LCSRs, 
analytically continuing integrands to the complex plane.
While this method was used as a check in \cite{DKMMO}, here we would like to emphasis its features and  
possible advantages over the traditional way of calculating sum rule amplitudes, especially when one performs
NLO calculations. 

\begin{figure}
\includegraphics[width=13cm]{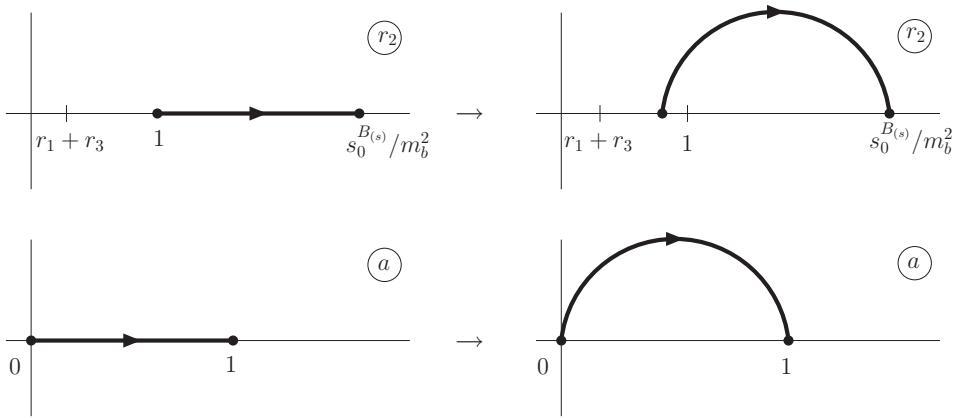}
\caption{\it Replacing the integration
intervals by the contours in the complex planes
of $a = u, X, X-\xi$ and $r_2 = s/m_b^2$ variables in the procedure
of numerical integration of LCSR amplitudes.
\label{fig:contour}}
\end{figure}
The main idea of the method is to deform the path of integration in order to
avoid poles which are located near real axes. Because of the Cauchy theorem, the deformation is legitimate if the integrand
is an analytic function inside the region bounded by the original and the new
path of integration. To check the analyticity of the integrand we have to examine its pole structure. 
In the case of NLO calculations, it is also necessary to examine
the position of cuts in logarithms and dilogarithms. Fortunately, 
there are just a few characteristic structures which have to be investigated. 

At LO, Eqs. (\ref{eq:corrF}-\ref{eq:corrFtens}), there are two possibilities to hit the pole, when
\begin{equation}
 m_b^2-(q+a p)^2 =0,\quad \mbox{ or} \quad  q\cdot p+a p^2 =0,
\label{dupli1}
\end{equation}
condition is fulfilled. In above, $a = {u,X,X-\xi}$ represents fraction of momenta between 0 and 1, over which it has to be 
integrated. 
For further considerations, it is convenient to introduce the notations 
\begin{eqnarray}
& & r_1=\frac{q^2}{m_b^2},~r_2=\frac{s}{m_b^2},~r_3=\frac{p^2}{m_b^2}=
\frac{m_K^2}{m_b^2}, \nonumber \\
& & \rho=(1-a)\,r_1+a\,r_2-a(1-a)\,r_3.
\label{dupli2}
\end{eqnarray}
By using (\ref{dupli2}), the conditions from (\ref{dupli1}) can be
written as
\begin{equation}
 m_b^2(1-\rho ) =0,\quad \mbox{or} \quad  \frac{m_b^2}{2}(r_2-r_1-r_3(1-2 a)) =0.
\label{dupli3}
\end{equation}
In the case of interest $r_2 >  r_1 +r_3 > 0$, the second condition from above
cannot be fulfilled and therefore there is no pole for $q\cdot p+a p^2 =0$. From the first condition in (\ref{dupli3}), it follows that
the integrand is approaching a pole when $\rho \to 1$. That happens for the real values
of $r_2$ and $a$ in the integration range.
Note that in Eqs. (\ref{eq:corrF}-\ref{eq:corrFtens},\ref{dupli1}) we have omitted an infinitesimal imaginary quantity $\rm{i}
\epsilon$ which appears in Feynman propagators.
Taking it into account, the exact position of the pole is given by the equation
$1-\rho -\rm{i}\epsilon =0$, which means that the poles are not located on the real axes
of $r_2$, but slightly below. As a consequence, one can deform the $r_2$ path of
integration into the upper half of the $r_2$ complex plane to avoid passing near 
the poles. If poles are far away from the integration path, the integration is numerically
completely stabile. The problem remains only when the poles coincide with the end
points of the integration. For the integration over $r_2$, the end points are at $1$ and
$s_0^{B_{(s)}}/m_b^2$. Then $\rho$ becomes
\begin{eqnarray}
\rho &=& a+(1-a)\,r_1-a(1-a)\,r_3  \quad \mbox{for} \quad r_2 =1, 
\label{dupli4}
\\
\rho &=& a\frac{s_0}{m_b^2}+(1-a)\,r_1-a(1-a)\,r_3 \quad \mbox{for} \quad
r_2=\frac{s_0^{B_{(s)}}}{m_b^2}>1. 
\label{dupli5}
\end{eqnarray}
In both cases $\rho$ can be equal to 1 in the range of integration over $a$.
In the case (\ref{dupli4}), $\rho$ is equal to 1 for $a=1$, which is the worst
possible case because this pole is located at the end point of two integrations.
In (\ref{dupli5}), $\rho=1$ for $0<a<1$, where, due to the specific values of $r_1$, $r_3$, and
$s_0$, $a$ cannot be near 0 or 1.
However, in both cases it is possible to move away from the poles.
The complete procedure is going in this way.
The first step is to shift the lower limit of $r_2$ (i.e., $s$) integration to any
point between $r_1+r_3<r_2<1$. That is legitimate because all integrands are real
for $r_2<1$ and we are interested only in an imaginary part of the integrand, as can be seen from
(\ref{eq:dispE}). The lower limit $r_1+r_3$ is necessary to evade the possibility to fulfill the
second condition from Eq. (\ref{dupli3}).
Now we move the operation of taking the imaginary part in (\ref{eq:dispE}) outside the integral.
As the third step we deform the path of the $r_2$ integration into the upper half
of the complex $r_2$-plane, so that all poles are away from the integration region.
For the calculation presented here, the new integration path is the semicircle in
the complex $r_2$-plane; see Fig.3. As mentioned before, the pole condition still can be
satisfied at the end points of integration.
However, since the lower end point of the $r_2$ integration is now $<1$, the pole condition
($\rho=1$) cannot be fulfilled at that end point. For the upper end point ($r_2= s_0^{B_{(s)}}/m_b^2$) 
the situation is as presented by Eq.(\ref{dupli5}). Because of the fact that
this pole is in the middle of the range of $a$ integration, it is possible to avoid it
now by deforming the contour of $a$-integration into the upper half of the complex
$a$-plane. Here, we again deform the integration path in the shape of
the semicircle, as shown in Fig.3. After that, all poles are away from the integration regions and all
integrals can be performed numerically without facing instabilities in the integration. At the end, 
it remains to take the imaginary part to get the final result.

For the NLO calculation, in addition, one has
to check analytical properties of appearing logarithms and polylogarithms.  It
happens that for the case of the interest in one of the logarithms it is impossible to
avoid crossing the cut when both variables $r_2$ and $a$ are continued to the upper
half of the complex space. To avoid crossing the cut, we have continued only $r_2$ to the complex plane. 
But now, the path of the integration for the variable $a$ will pass near the pole
when $r_2$ is approaching the endpoint $s_0^{B_{(s)}}/m_b^2$. 
Although the problem can be
cured by the variable transformation and a sophisticated analytical continuation, considering the precision
needed for the calculation, such a sophisticated method is obsolete indeed. 
The numerical instability shows up in the third significant digit, and therefore does not affect 
the final numerical results. 

\section{Updated predictions for the $B_{(s)}\to K$   form factors}

All input parameters are listed in Appendix B. 
It is a compilation of the most recent determination of parameters entering the calculation. 

The renormalization scale is given by the expression $\mu_{(s)} = \sqrt{m_{B_{(s)}}^2 - m_b^2}$. Therefore, for 
the $f_{BK}^{0,+,T}$ form factors we use $\mu= 3$ GeV and for $f_{B_sK}^{0,+,T}$ the renormalization scale is $\mu_s= 3.4$ GeV. 
As usual, we will check the sensitivity of the results on the variation of above scales and will include it in the error estimation. 

From the general LCSR expressions for the form factors, (\ref{eq:fplusLCSR}-\ref{eq:fTLCSR}), one can note that the decay constant 
$f_{B}$ (and correspondingly $f_{B_s}$ for $B_s \to K$ decays) enters the calculation. To reduce the dependence of the form factors on 
the input parameters, we replace $f_B$ and $f_{B_s}$ by two-point sum rule expressions 
in the $\overline{MS}$ scheme from \cite{JL} to $O(\alpha_s, m_s^2)$ accuracy and calculate them for our preferred values of parameters. 

\begin{figure}[b]
\begin{center}
\includegraphics[width=8.5cm]{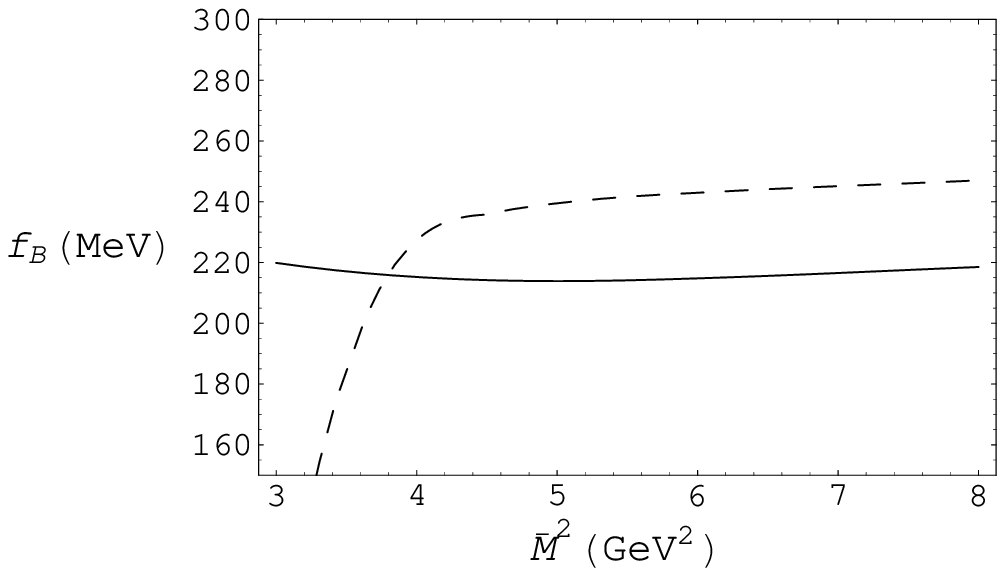}
\hfill
\includegraphics[width=8.5cm]{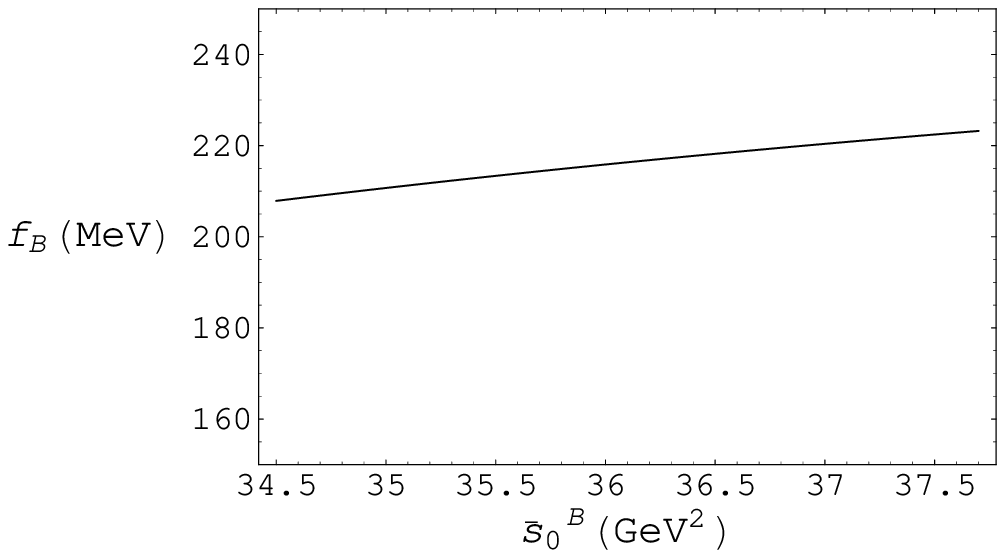}
\\
\hspace*{2cm} (a) \hspace*{9cm} (b)
\caption{ \it Dependence of $f_B$ on (a) the Borel parameter $\overline{M}^2$ shown for $\mu = 3$ GeV (solid line) and 
$\mu = 6$ GeV (dashed line) and (b) the effective threshold parameter $\overline{s}_0^B$ using the central values of all other input parameters.} 
\label{fig-fB}
\end{center}
\end{figure}

\begin{figure}[h]
\begin{center}
\includegraphics[width=8.5cm]{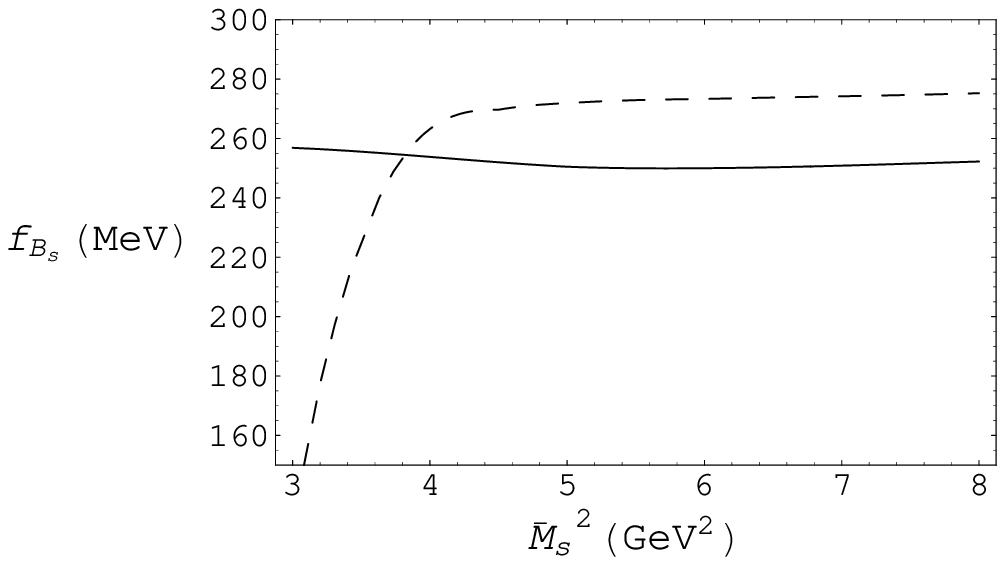}
\hfill
\includegraphics[width=8.5cm]{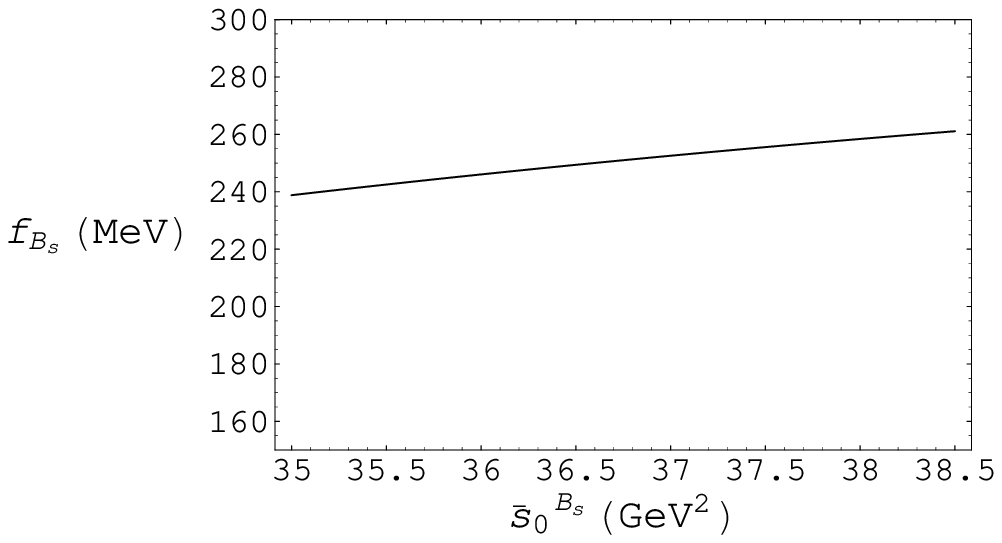}
\\
\hspace*{2cm} (a) \hspace*{9cm} (b)
\caption{ \it Dependence of $f_{B_s}$ on (a) the Borel parameter $\overline{M}_s^2$ shown for $\mu = 3.4$ GeV (solid line) and 
$\mu = 6$ GeV (dashed line) and (b) the effective threshold parameter $\overline{s}_0^{B_s}$ using the central values of all other input parameters.} 
\label{fig-fBs}
\end{center}
\end{figure}

The usual method for deriving the working region of Borel parameters and determining effective threshold parameters is used. 
We investigate the behavior of the perturbative expansion and smallness of the continuum contribution (to be less then $30\%$ of the total 
contribution), and require that the derivative over the Borel parameter of the expression for a particular decay constant,  
which gives the sum rule for $m_B^2$ ($m_{B_s}^2$),  does not deviate more than $0.5-1 \%$ from the experimental values for those masses. 
We obtain  the following sets of parameters: $\overline{M}^2 = 5 \pm 1 \, {\rm GeV^2}$ and
 $\overline{s}_0^B =35.6^{-0.9}_{+2.1}\, {\rm GeV^2}$ for the 
$B$-meson decay constant $f_B$ calculated at $\mu = 3$ GeV, and 
$\overline{M}_{s}^2 = 6.1 \pm 1.5\,{\rm GeV^2}$ and 
$\overline{s}_0^{B_s} = 36.6^{-1.6}_{+1.9}\, {\rm GeV^2}$ for $f_{B_s}$ calculated at $\mu_s = 3.4$ GeV. 
Note that the calculated central values of $\overline{s}_0^{B_{(s)}}$ follow the naive relation $\overline{s}_0^{B_s} - \overline{s}_0^{B} 
\simeq m_{B_s}^2 - m_B^2 \simeq 1 \, {\rm  GeV^2}$. 
Employing these values, the resulting decay constants are
\be
f_B = 214 \pm 18 \; {\rm MeV} , \,  f_{B_s} =  250 \pm 20 \; {\rm MeV}. 
\ee
In the LCSR expression for the form factors some of the uncertainties are going to cancel in the ratios, and therefore the error 
intervals of the $f_{B}$ and $f_{B_s}$ input will reduce, as one can see from the following numbers, 
$f_B = 214 \pm 9\; {\rm MeV}$ and $f_{B_s} =  250 \pm 11\; {\rm MeV}$, where the calculated error intervals 
come from the variation of $\overline{s}_0^{B_{(s)}}$ and $\overline{M}_{(s)}^2$ only. 
The dependence of the decay constants on $\overline{M}_{(s)}^2$ and $\overline{s}_0^{B_{(s)}}$ appears to be mild, 
as shown in Figs. \ref{fig-fB} and \ref{fig-fBs}. 
%
The SU(3) violation among decay constants is \cite{JL}  
\be
\frac{f_{B_s}}{f_B} = 1.16 \pm 0.05\, , 
\ee 
which nicely agrees with the values obtained from the lattice calculation and by different quark models \cite{O,A,IKKR}.  

\begin{figure}[h]
\begin{center}
\includegraphics[width=8.5cm]{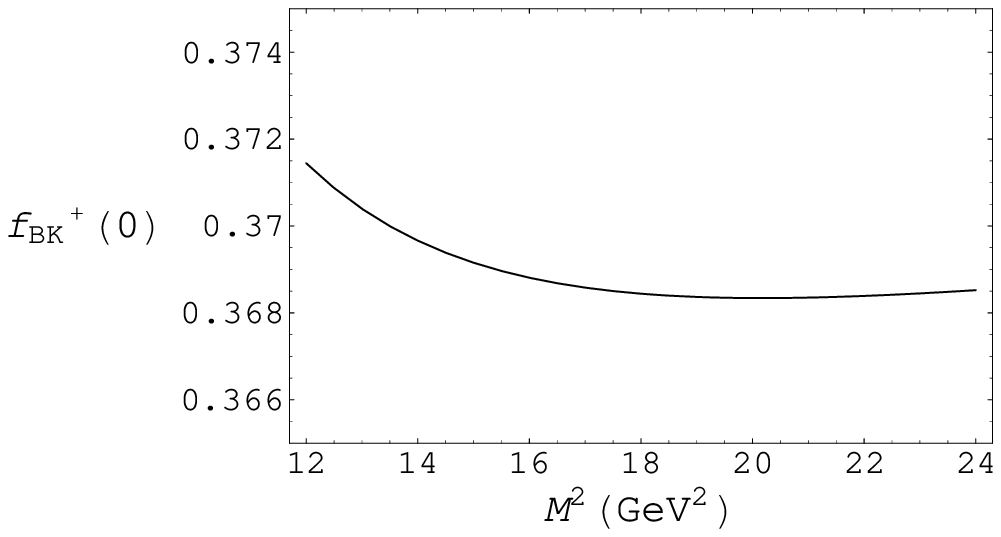}
\hfill
\includegraphics[width=8.5cm]{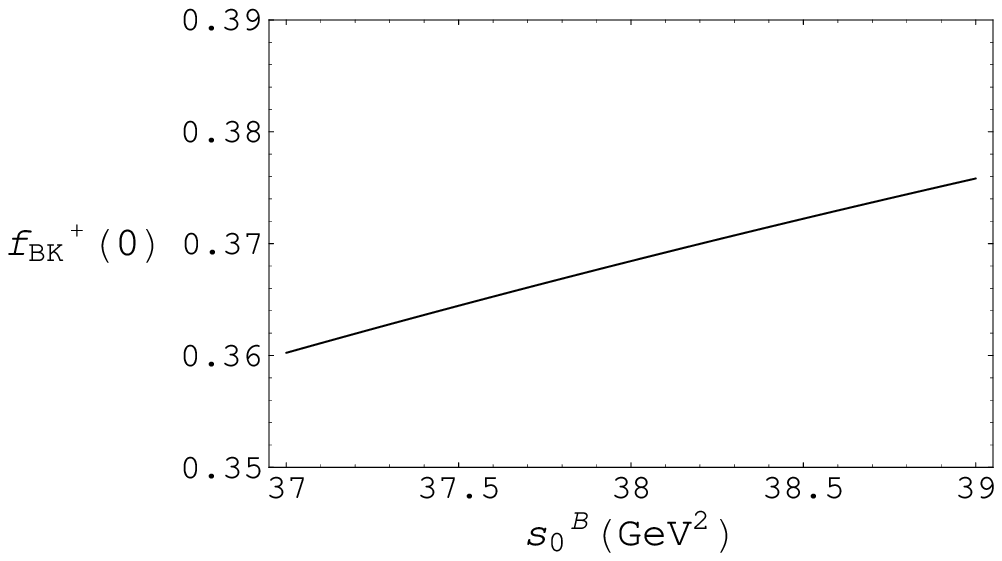}
\\
\hspace*{2cm} (a) \hspace*{9cm} (b)
\caption{ \it Dependence of $f_{BK}^+$ on (a) the Borel parameter $M^2$ and
(b) the effective threshold parameter $s_0^B$ using the central values of all other input parameters.}
\label{fig-fBK}
\end{center}
\end{figure}

\begin{figure}[h]
\begin{center}
\includegraphics[width=8.5cm]{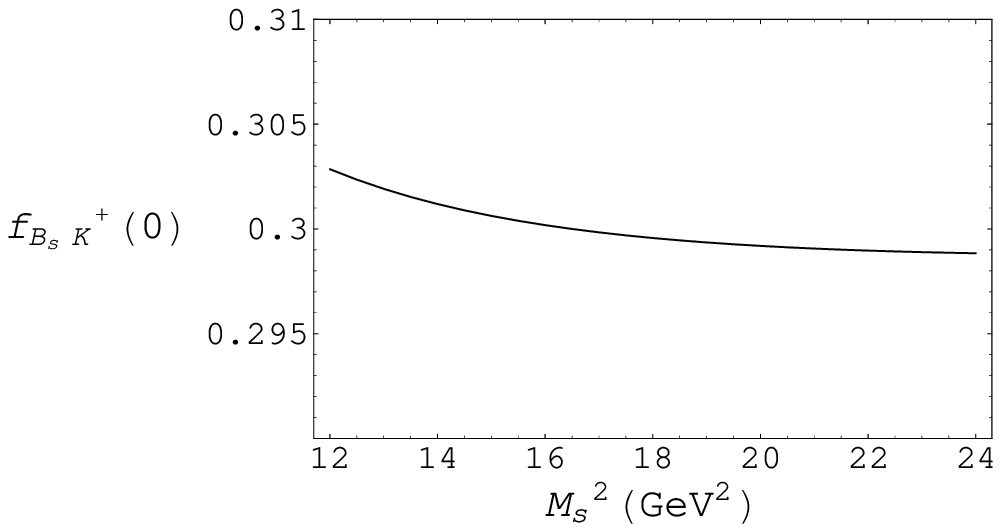}
\hfill
\includegraphics[width=8.5cm]{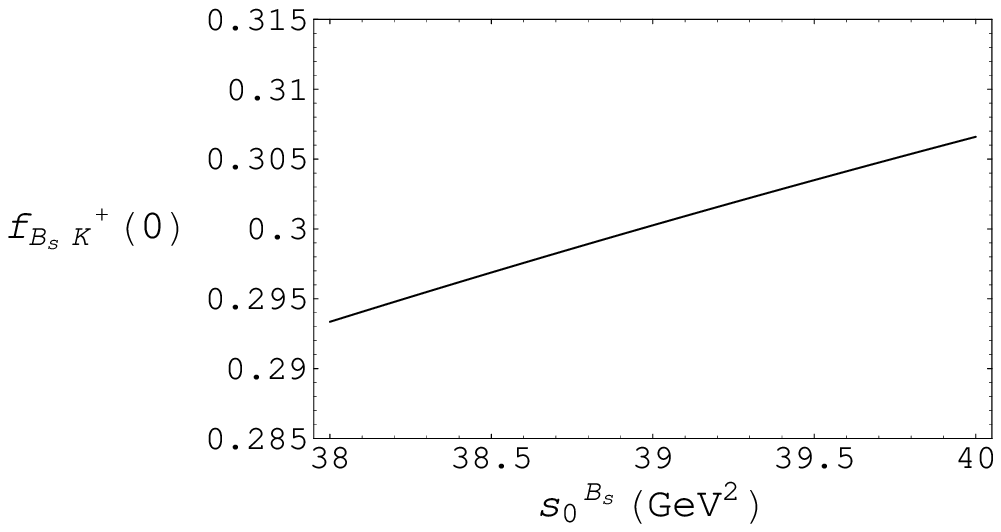}
\\
\hspace*{2cm} (a) \hspace*{9cm} (b)
\caption{ \it Dependence of $f_{B_sK}^+$ on (a) the Borel parameter $M_s^2$ and (b) the effective threshold parameter 
$s_0^{B_s}$ using the central values of all other input parameters.}
\label{fig-fBsK}
\end{center}
\end{figure}

The method of extraction of the Borel parameters $M$ and $M_s$, and the effective thresholds $s_0^B$ and $s_0^{B_s}$ for 
$f_{B_{(s)}K}^{+,0,T}$ form factors is similar to the above, and it is the same as described 
in \cite{DKMMO}. We require that the subleading twist-4 terms in the LO are small,
less than $3\%$ of the LO twist-2 term, that
the NLO corrections of twist-2 and twist-3 parts are not exceeding $30\%$ of their LO counterparts, and that the subtracted continuum remains
small, which fixes the allowed range of $M^2_{(s)}$. The effective threshold parameters are again fitted so that 
the derivative over $-1/M_{(s)}^2$ of the expression of the complete LCSRs for a particular form factor 
reproduces the physical masses $m_{B_{(s)}}^2$ with a high accuracy of $O(0.5 \%-1 \%)$ in the stability region of the sum rules. 
These demands provide us the following central values for the sum rule parameters: $M^2 = 18\,{\rm GeV}^2$, $s_0^B = 38 \,{\rm GeV}^2$, 
$M_s^2 = 19\,{\rm GeV}^2$, and $s_0^{B_s} = 39 \,{\rm GeV}^2$. The dependence of the form factors on the these parameters is 
depicted in Figs. \ref{fig-fBK} and \ref{fig-fBsK}. 
%
 
The complete numerical analysis yields the following
predictions for the vector $B\to K$ and $B_s \to K$ form factors at zero momentum transfer:
\be
f^+_{BK}(0)= 0.368 \pm 0.011\bigg|_{a_1,a_2} \pm 0.008\bigg|_{M,\overline{M}}
\,_{-0.008}^{+0.017}\bigg|_{\mu} \pm 0.006\bigg|_{m_b} 
\,^{+0.036}_{-0.024}\bigg|_{\mu_\pi}+ 0.026\bigg|_{m_K^2 \, {\rm \, at \, NLO}}\,,
\label{eq:fBKres}
\ee
\be
f^+_{B_sK}(0)= 0.300 \pm 0.007\bigg|_{a_1,a_2} \,_{-0.007}^{+0.006}\bigg|_{M_s,\overline{M}_s}
\,_{-0.003}^{+0.004}\bigg|_{\mu} \,_{-0.002}^{+0.001}\bigg|_{m_b} 
\,^{+0.034}_{-0.020}\bigg|_{\mu_\pi}+ 0.026\bigg|_{m_K^2 \, {\rm \, at \, NLO}}\,,
\label{eq:fBsKres}
\ee
where the central value for $f^+_{BK}$ is calculated at
$\mu=3.0$ GeV, $M^2=18.0$ GeV$^2$, $s_0^B=38$ GeV$^2$,
$\overline{M}^2=5.0$ GeV$^2$, and $\overline{s}_0^B=35.6$ GeV$^2$, and for $f^+_{B_sK}$ at  
$\mu_s=3.4$ GeV, $M^2_s=19.0$ GeV$^2$, $s_0^{B_s}=39$ GeV$^2$, $\overline{M}_s^2=6.1$ GeV$^2$, $\overline{s}_0^{B_s}=36.6$ GeV$^2$. 
The central values for the parameters of the twist-2 kaon DA are $a_1^K(1 \mbox{GeV})=0.10$  and $a_2^K(1 \mbox{GeV})=0.25$. 
The last error in (\ref{eq:fBKres}) and (\ref{eq:fBsKres}) comes from the neglected $O(m_K^2)$ effects at NLO. 

\begin{figure}[t]
\begin{center}
\includegraphics[width=9.5cm]{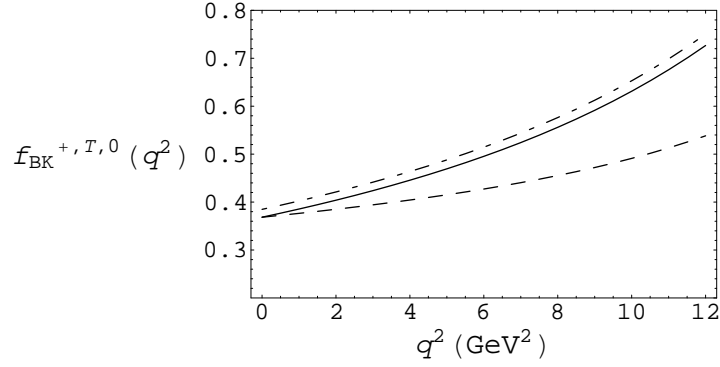}
\caption{ \it The LCSR prediction for
form factors $f^+_{BK}(q^2)$ (solid line), $f^0_{BK}(q^2)$ (dashed line),
and $f^T_{BK}(q^2)$ (dash-dotted line)
at $0<q^2<12$ GeV$^2$ and for $\mu = 3$ GeV, $s_0^{B} = 38\,{\rm GeV}^2$, $M^2 = 18 \, {\rm GeV}^2$, and the central values of all 
other input parameters. }
\label{fig-fK}
\end{center}
\end{figure}

\begin{figure}[t]
\begin{center}
\includegraphics[width=9.5cm]{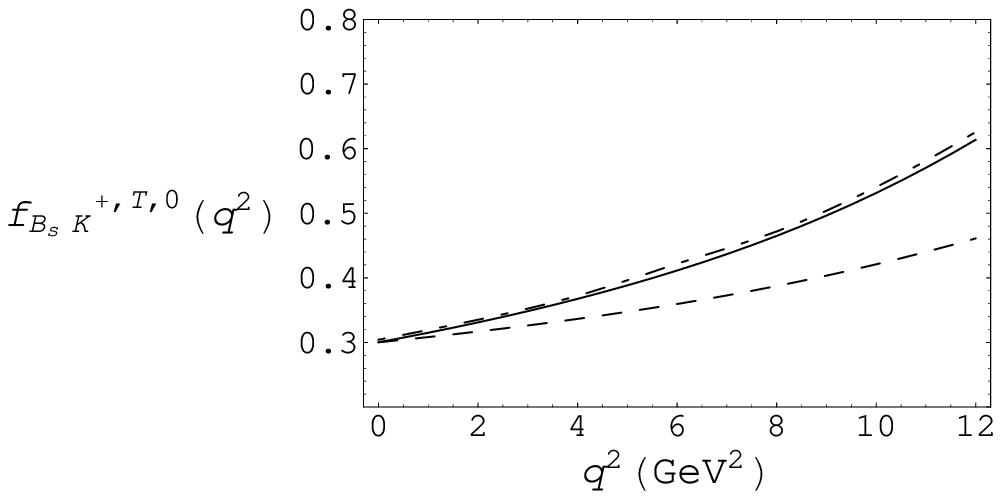}
\caption{ \it The LCSR prediction for
form factors $f^+_{B_sK}(q^2)$ (solid line), $f^0_{B_sK}(q^2)$ (dashed line),
and $f^T_{B_sK}(q^2)$ (dash-dotted line)
at $0<q^2<12$ GeV$^2$ and for $\mu = 3.4$ GeV, $s_0^{B_s} = 39\,{\rm GeV}^2$, $M_s^2 = 19 \, {\rm GeV}^2$, and the central values of all other 
input parameters. }
\label{fig-fKs}
\end{center}
\end{figure}

Finally, adding all uncertainties in quadratures, and to be on the safe side, allowing that the real mass corrections at NLO could reduce the final 
result, we obtain the following values 
for different $B \to K$ form factors:
\ba
f_{BK}^+(0) = f_{BK}^{0}(0)  = 0.36^{+0.05}_{-0.04}\,, 
\\
f_{BK}^T(0) = 0.38\pm 0.05\,,  
\label{eq:res}
\ea
and for $B_s\to K$ form factors, 
\ba
f_{B_sK}^+(0) = f_{B_sK}^{0}(0)  = 0.30^{+0.04}_{-0.03}\,,
\\
f_{B_sK}^T(0) = 0.30\pm 0.05 \, . 
\label{eq:resS}
\ea
Their $q$ dependence is shown in Figs.\ref{fig-fK} and Fig.\ref{fig-fKs}, where the values in the allowed LCSR kinematical regime are shown. 
The above results for the form factors are in an overall agreement with those extracted by other 
methods \cite{A,IKKR,MS}.
The predictions include also the uncertainty from the inclusion of $m_K^2$ effects at NLO, which is 
relatively large, as can be deduced from (\ref{eq:fBKres}) and (\ref{eq:fBsKres}). However, one has to be aware that this 
error only gives us a flavor of the size of neglected mass corrections at NLO, since the $m_s$ effects in the hard-scattering 
amplitude could not be included, and we expect 
that there will be a partial cancellation among $m_s$ and $m_K^2$ contributions, being of similar size. 
At the leading order, the inclusion of $m_s$ effects (appearing only in DAs) and $m_K^2$ effects reduces the results by $2.5\%-4\%$,
 depending of the value of $q^2$. 

In order to be able to comment on the SU(3)-breaking effects, 
these values have to be compared with the results obtained by the same method for 
$B \to \pi$ form factors \cite{DKMMO}, which we quote here: 
\ba
f_{B\pi}^+(0) = f_{B\pi}^{0}(0)  = 0.26^{+0.04}_{-0.03}\,,
\\
f_{B\pi}^T(0) = 0.255 \pm 0.035\,.
\ea

By varying parameters in a correlated way, finally  we predict the following SU(3)-breaking ratios:
\ba
\frac{f_{BK}^+(0)}{f_{B\pi}^+(0)} &=& 1.38^{+0.11}_{-0.10}\,,  \qquad\qquad\qquad \frac{f_{B_sK}^+(0)}{f_{B\pi}^+(0)} = 1.15^{+0.17}_{-0.09}\,, 
\label{eq:Bpiratio}
\\
\frac{f_{BK}^T(0)}{f_{B\pi}^T(0)} &=& 1.49^{+0.18}_{-0.06}\,,  \qquad\qquad\qquad \frac{f_{B_sK}^T(0)}{f_{B\pi}^T(0)} = 1.17^{+0.15}_{-0.11}\,.
\label{eq:BpiTratio}
\ea
The complete SU(3) violation comes from SU(3)-breaking corrections in all parameters, mainly from $f_K/f_{\pi}$, $\mu_{K}/\mu_{\pi}$, 
from the difference in the sum rule parameters, $s_B$ and $M$, as well as from the difference in the  $f_{B_s}$ and $f_B$ ratio. 

Compared with the values from 
the second paper in \cite{KMM}, where the similar LCSR analysis was done, we find nice agreement with the results in (\ref{eq:Bpiratio}). 

It is also interesting to explore an overall SU(3)-breaking factor, which appears in factorization models for $B_{(s)} \to  K \pi$ 
amplitudes \cite{CGR,Gronau}: 
\be
\xi = \frac{f_K}{f_{\pi}}\frac{f^+_{B\pi}(m_K^2)}{f_{B_sK}^+ (m_{\pi}^2)}\frac{m_B^2 - m_{\pi}^2}{m_{B_s}^2 - m_K^2} = 
1.01^{+0.07}_{-0.15}. 
\ee
For the $f_K/f_{\pi}$ ratio we use (\ref{eq:ratio}). 
Although there is a SU(3) violation among form factors and in the masses, the predicted value for $\xi$ 
shows almost exact SU(3) symmetry. On the other hand, the above ratio enters the prediction for $B_s \to K^- \pi^+$ amplitude obtained by employing 
U-spin symmetry.  U-spin symmetry cannot be trusted \cite{KMM}, as we can note by inspecting another U-spin relation. By neglecting penguin and 
annihilation contributions, under the U-spin symmetry assumption 
$A_{fact}(B_s \to K^+K^-)/A_{fact}(B_d \to \pi^+ \pi^-) \sim 1$, \cite{Gronau,F,KMM},  while our prediction amounts to
\be
\frac{A_{fact}(B_s \to K^+K^-)}{A_{fact}(B_d \to \pi^+ \pi^-) } 
= \frac{f_K}{f_{\pi}} \frac{f_{B_sK}^+ (m_{K}^2)}{f^+_{B\pi}(m_{\pi}^2)} \frac{m_{B_s}^2 - m_K^2}{m_B^2 - m_{\pi}^2} = 
1.41^{+0.20}_{-0.11} \, , 
\ee
a quite substantial U-spin violation. 

\section{Summary}

In this paper we have investigated the SU(3)-symmetry breaking effects in the $B \to K$ and $B_s \to K$ form factors. The analysis has involved 
the SU(3)-breaking corrections both in the LO (up to twist-4 corrections), as well as in the NLO calculation, estimating SU(3) corrections 
for the twist-2 and twist-3 contributions. 
Although at NLO we were not able to consistently include $O(m_s) \sim O(m_K^2)$ effects, we have included $m_K^2$ effects in the 
error analysis of our results. 
We have presented a method of numerical integration of sum rule amplitudes, which greatly facilitates the calculation, especially 
the calculation of the radiative corrections. 
By investigating some of the SU(3) and U-spin symmetry relations, we have shown that such relations have to be considered with a precaution, 
since some of them can be badly broken.

\section*{Acknowledgments}
We are grateful to A. Khodjamirian, A. Lenz, and K. Passek-Kumeri\v cki 
for useful discussions. 
The work is supported by the
Ministry of Science, Education, and Sport of the 
Republic of Croatia, under Contract No. 098-0982930-2864 and by the Alexander von Humboldt Foundation.

\appendix

\section{Explicit formulas for the leading-order LCSR expressions}

Although the intention of this paper is to promote the numerical method for calculating LCSR amplitudes, for which the explicit 
expressions for the imaginary parts are superfluous, because the result can be obtained by direct integration of the starting 
amplitudes (at LO they are given by Eqs.(\ref{eq:corrF}-\ref{eq:corrFtens})), we have decided to list here the 
LO LCSR expressions for $f_{B_{(s)}K}$ form factors, since 
to our best knowledge, these expressions were never clearly presented in a form which includes complete mass corrections. 
 
The LO part of the $f_{BK}^+$ LCSR, (\ref{eq:fplusLCSR}), has the following form:
\ba
&& F_0(q^2,M^2,s_0^B)= m_b^2f_K\int\limits_{u_0}^1 du\,
e^{-\frac{m_b^2-q^2\bar{u} + m_K^2 u \bar{u}}{u M^2}} 
\Bigg\{\frac{\varphi_K(u)}{u}\qquad\qquad\qquad\qquad
\nonumber
\\
&& \qquad +\frac{\mu_K}{m_b}\Bigg[ \phi_{3K}^p(u)
+\frac{1}{6}\Bigg( 2 \frac{\phi_{3K}^\sigma(u)}{u}
-\frac{1}{m_b^2 -q^2 + u^2 m_K^2} 
\bigg ( ( m_b^2 +q^2 - u^2 m_K^2 ) \frac{d \phi_{3K}^\sigma(u)}{du} 
\nonumber \\
&& \qquad \quad - \frac{4 u m_K^2 m_b^2}{m_b^2 -q^2 + u^2 m_K^2}  \phi_{3K}^\sigma(u) \bigg )\Bigg)\Bigg]
\nonumber 
\\
&& \qquad +\frac{1}{m_b^2-q^2 + u^2 m_K^2}
\Bigg[ 
u\psi_{4K}(u)+ \left ( 1 - \frac{2 u^2 m_K^2}{m_b^2-q^2 + u^2 m_K^2} \right )\int\limits_0^u dv \psi_{4K}(v)
\nonumber 
\\
&& \qquad  
- \frac{m_b^2}{4}\frac{u}{m_b^2-q^2 + u^2 m_K^2} 
\left ( \frac{d^2}{du^2} - \frac{6 u m_K^2}{m_b^2-q^2 + u^2 m_K^2} \frac{d}{du} + 
\frac{12 u m_K^4}{(m_b^2-q^2 + u^2 m_K^2)^2} \right ) \phi_{4K}(u) 
\nonumber 
\\
&& \qquad - \left ( \frac{d}{du} - \frac{2 u m_K^2}{m_b^2-q^2 + u^2 m_K^2} \right ) 
\left ( \left (\frac{f_{3K}}{m_b f_K} \right ) I_{3K}(u) + I_{4K}(u)  - \frac{d I_{4K}^{\Xi}(u)}{du}  \right )
\nonumber \\
&& \qquad
-\frac{ 2 u m_K^2 }{m_b^2-q^2 + u^2 m_K^2} 
\left ( u\frac{d}{du} + \left (1- \frac{4 u^2 m_K^2}{m_b^2-q^2 + u^2 m_K^2} \right ) \right ) \overline{I}_{4K}(u) 
\nonumber 
\\
&& \qquad + \frac{ 2 u m_K^2 (m_b^2-q^2 - u^2 m_K^2)}{(m_b^2-q^2 + u^2 m_K^2)^2} 
\left ( \frac{d}{du} - \frac{6 u m_K^2}{m_b^2-q^2 + u^2 m_K^2} \right ) \int_u^1 d\xi \overline{I}_{4K}(\xi)
\Bigg]
\Bigg\}
\nonumber \\
&& \qquad +  \frac{m_b^2f_K e^{-\frac{m_b^2}{ M^2}}}{m_b^2 -q^2 +  m_K^2}\bigg [ - \int_0^1 dv \psi_{4K}^{WW}(v) 
+ \frac{m_b^2}{4}\frac{1}{m_b^2-q^2 + m_K^2} \bigg ( \frac{d \phi_{4K}^{WW}(u)}{du} \bigg )_{u \to 1} \bigg ] \, , 
\nonumber \\
\label{eq:fplusBpiLCSRcontrib}
\ea
where $\bar{u}=1-u$, $u_0=\left (q^2 - s_0^B + m_K^2 + \sqrt{(q^2 - s_0^B + m_K^2)^2 - 4 m_K^2 (q^2 - m_b^2)} \right )/(2 m_K^2) $,  and the 
short-hand notations
introduced for the integrals over three-particle DA's are
\ba
&& I_{3K}(u)=\int\limits_0^u \!d\alpha_1\!\!\!
\int\limits_{(u-\alpha_1)/(1-\alpha_1)}^1\!\!\!\!\! \frac{dv}{v} \,\,
\left [ 4 v p \cdot q - (1 - 2v) u m_K^2 \right ] \Phi_{3K}(\alpha_i)
\Bigg|_{\begin{array}{l}
\alpha_2=1-\alpha_1-\alpha_3,\\
\alpha_3=(u-\alpha_1)/v
\end{array}
}
\,,
\nonumber
\\
&& I_{4K}(u)=\int\limits_0^u\! d\alpha_1\!\!\!
\int\limits_{(u-\alpha_1)/(1-\alpha_1)}^1\!\!\!\!\! \frac{dv}{v} \,\,
\Bigg[ 2 \Psi_{4K}(\alpha_i)-  \Phi_{4K}(\alpha_i) + 2 \widetilde{\Psi}_{4K}(\alpha_i) - \widetilde{\Phi}_{4K}(\alpha_i) 
\Bigg]
\Bigg|_{\begin{array}{l}
\alpha_2=1-\alpha_1-\alpha_3,\\
\alpha_3=(u-\alpha_1)/v
\end{array}
}
\,.
\nonumber
\\
&& \overline{I}_{4K}(u)=\int\limits_0^u\! d\alpha_1\!\!\!
\int\limits_{(u-\alpha_1)/(1-\alpha_1)}^1\!\!\!\!\! \frac{dv}{v} \,\,
\Bigg[
\Psi_{4K}(\alpha_i) + \Phi_{4K}(\alpha_i) + 
 \widetilde{\Psi}_{4K}(\alpha_i)+ \widetilde{\Phi}_{4K}(\alpha_i) 
\Bigg]
\Bigg|_{\begin{array}{l}
\alpha_2=1-\alpha_1-\alpha_3,\\
\alpha_3=(u-\alpha_1)/v
\end{array}
}
\,.
\nonumber
\\
&& I_{4K}^{\Xi}(u)= \int\limits_0^u \! d\alpha_1\!\!\!
\int\limits_{(u-\alpha_1)/(1-\alpha_1)}^1\!\!\!\!\! \frac{dv}{v} \,\,
\Bigg[
v (1-v) \Xi_{4K}(\alpha_i)
\Bigg]
\Bigg|_{\begin{array}{l}
\alpha_2=1-\alpha_1-\alpha_3,\\
\alpha_3=(u-\alpha_1)/v
\end{array}
}
\,.
\label{eq:fplusBpiLCSR3part}
\ea
The twist-4 two-particle DAs are defined with the help of the two DAs \cite{BBL}: $\phi_{4K} = \phi_{4K}^{T4} + \phi_{4K}^{WW}$ and 
$\psi_{4K} = \psi_{4K}^{T4} + \psi_{4K}^{WW}$. 
\\
The leading-order LCSR 
for $f^+_{BK} + f^-_{BK}$, (\ref{eq:fplminLCSR}), looks like
\ba
&& \widetilde{F}_0(q^2,M^2,s_0^B)= 
m_b^2 f_K\int\limits_{u_0}^1 du\, e^{-\frac{m_b^2-q^2\bar{u} + m_K^2 u \bar{u}}{uM^2}}
\Bigg\{\frac{\mu_K}{m_b}\Bigg(\frac{\phi_{3K}^p(u)}{u}
+\frac{1}{6u}\frac{d\phi_{3K}^{\sigma}(u)}{du}\Bigg)
\nonumber
\\
&& \qquad +\frac{1}{m_b^2-q^2 + u^2 m_K^2} 
\Bigg [ \psi_{4K}(u) - \frac{2 u m_K^2}{m_b^2 - q^2 + u^2 m_K^2} \int_0^u dv \psi_{4K}(v) 
\nonumber \\
&& \qquad + m_K^2 \left (\frac{d}{du} - \frac{2 u m_K^2}{m_b^2 - q^2 + u^2 m_K^2} \right ) 
\left (\frac{f_{3K}}{f_K m_b} \right ) \widetilde{I}_{3K}(u) 
\nonumber \\
&& \qquad + \frac{2 u m_K^2}{m_b^2 - q^2 + u^2 m_K^2} \bigg ( \frac{d^2}{du^2} - \frac{6 u m_K^2}{m_b^2 - q^2 + u^2 m_K^2} \frac{d}{du} 
+ \frac{12 u^2 m_K^4}{(m_b^2 - q^2 + u^2 m_K^2)^2} \bigg ) \int_u^1 d\xi \overline{I}_{4K}(\xi)
\Bigg ]
\Bigg\}
\nonumber \\
&& \qquad +  \frac{m_b^2f_K e^{-\frac{m_b^2}{ M^2}}}{m_b^2 -q^2 +  m_K^2}\bigg [ - \int_0^1 dv \psi_{4K}^{WW}(v) \bigg ] \, 
\,.
\label{eq:fplminBpiLCSRcontrib}
\ea
where 
\ba
&& \widetilde{I}_{3K}(u)=\int\limits_0^u \!d\alpha_1\!\!\!
\int\limits_{(u-\alpha_1)/(1-\alpha_1)}^1\!\!\!\!\! \frac{dv}{v} \,\,
\left [ (3 - 2v) \right ] \Phi_{3K}(\alpha_i)
\Bigg|_{\begin{array}{l}
\alpha_2=1-\alpha_1-\alpha_3,\\
\alpha_3=(u-\alpha_1)/v
\end{array}
}
\,.
\ea
Finally, the leading-order LCSR for the penguin form factor, (\ref{eq:fTLCSR}), reads 
\ba
&& F_{0}^{T}(q^2,M^2,s_0^B)= m_b f_K\int\limits_{u_0}^1 
du\, e^{-\frac{m_b^2-q^2\bar{u} + m_K^2 u \bar{u}}{uM^2}} 
\Bigg\{\frac{\varphi_K(u)}{u}
\nonumber \\
&& \quad 
-\frac{m_b\mu_K}{3(m_b^2-q^2 + u^2 m_K^2)}
\left (\frac{d\phi_{3K}^\sigma(u)}{du} - \frac{2 u m_K^2}{m_b^2 - q^2 + u^2 m_K^2} \phi_{3K}^\sigma(u) \right )
\nonumber
\\
&& \quad
+\frac{1}{m_b^2-q^2 + u^2 m_K^2}
\Bigg[\left ( \frac{d}{du} - \frac{2 u m_K^2}{m_b^2 - q^2 + u^2 m_K^2} \right )\bigg ( \frac{1}{4} \phi_{4K}(u) - I_{4K}^T(u) 
+ \frac{d I_{4K}^{\Xi}(u) }{du} \bigg )
\nonumber \\
&& \quad 
- \frac{m_b^2\,u}{4(m_b^2-q^2 + u^2 m_K^2)}\bigg (\frac{d^2}{du^2} - \frac{6 u m_K^2}{m_b^2 - q^2 + u^2 m_K^2} 
\frac{d}{du} + \frac{12 u m_K^4}{(m_b^2 - q^2 + u^2 m_K^2)^2} \bigg ) \phi_{4K}(u)
\Bigg]\Bigg\}
\nonumber \\
&& +  \frac{m_b f_K e^{-\frac{m_b^2}{ M^2}}}{m_b^2 -q^2 +  m_K^2}\bigg [ 
\frac{m_b^2}{4}\frac{1}{m_b^2-q^2 + m_K^2} \bigg ( \frac{d \phi_{4K}^{WW}(u)}{du} \bigg )_{u \to 1} \bigg ]
\, 
\label{eq:fTBpiLCSRcontrib}
\ea
and
\ba
&I_{4K}^T(u)&= \int\limits_0^u\! d\alpha_1\!\!\!
\int\limits_{(u-\alpha_1)/(1-\alpha_1)}^1\!\!\!\!\! \frac{dv}{v} \,\,
\Bigg[2\Psi_{4K}(\alpha_i)-(1-2v)\Phi_{4K}(\alpha_i)
\nonumber
\\
&&  \qquad\qquad\qquad 
+ 2(1-2v)\widetilde{\Psi}_{4K}(\alpha_i)-
\widetilde{\Phi}_{4K}(\alpha_i)\Bigg]
\Bigg|_{\begin{array}{l}
\alpha_2=1-\alpha_1-\alpha_3,\\
\alpha_3=(u-\alpha_1)/v
\end{array}
}\,.
\label{eq:fTBpiLCSR3part}
\ea
Note the appearance of the surface terms in form factors above. Being proportional to the Wandzura-Wilczek part of 
$\phi_{4K}$ and $\psi_{4K}$, they vanish for $m_K \to 0$. 

The expressions for $f_{B_s K}^{+,0,T}$ form factors follows from above, by replacing $u$ by $1-u$ and by interchanging 
$\alpha_1$ and $\alpha_2$, and $m_d$ and $m_s$ in the kaon DAs, i.e., by replacing DAs of $\overline{K}^0$ by those for $K^0$. 

%
\begin{table}[t]
\begin{center}
\begin{tabular}{|c|c|}
\hline
Parameter & Value at $\mu=1$ GeV \\
\hline
$\overline{m}_s$ & $128 \pm 21$ MeV \\
\hline
$a_1^K$  &  $0.10 \pm 0.04$ \protect{\cite{CKP}}\\
$a_2^K$ & $0.25 \pm 0.15$ \\
$a_{> 2}^K$ & 0 \\
\hline
$f_{3K}$ & $0.0045\pm 0.0015$ GeV$^2$ \\
$\omega_{3K}$ & $-1.2\pm 0.7$ \\
$\lambda_{3K}$ & $1.6 \pm 0.4$ \\
\hline
$\delta^2_{K}$& $0.20\pm 0.06$  GeV$^2$ \\
$\omega_{4K}$ & $0.2\pm 0.1$ \\
$\kappa_{4K}$ & $-0.09 \pm 0.02$ \\
\hline
\end{tabular}
\\
\end{center}
\caption{\it Input parameters for the kaon DA's \protect{\cite{BBL,DKMMO}}.}
\label{tab-1}
\end{table}
\begin{table}[t]
\begin{center}
\begin{tabular}{|c|c|}
\hline
Parameter & Value  \\
\hline
$\alpha_s(m_Z)$ & $0.1176 \pm 0.002$ \\
\hline
$\langle \bar{q}q \rangle $(1 GeV) & $-(246^{+18}_{-19}\,{\rm MeV})^3$ \\
$\langle \bar{s}s \rangle $/$\langle \bar{q}q \rangle $ & $0.8 \pm 0.3$ \\
$\langle \alpha_s/\pi GG \rangle $ & $0.012^{+0.006}_{-0.012}\, {\rm GeV^4}$ \\
$m_0^2$ & $0.8 \pm 0.2\,{\rm  GeV}^2$
\\
\hline
\end{tabular}
\\
\end{center}
\caption{\it Additional input parameters for the $f_B$ and $f_{B_s}$ sum rules.}
\label{tab-2}
\end{table}

\begin{table}[t]
\begin{center}
\begin{tabular}{|l|l|l|l|l|}
\hline
$\phi_0^K = 0$ &  $\tilde{\phi}_0^K = -\frac{1}{3}\delta_K^2$ & $\psi_0^K = -\frac{1}{3}\delta_K^2$ &  $\theta_0^K = 0$ &
$\Xi_0^K = \frac{1}{5}\delta_K^2 a_1^K$ 
\\
$\phi_1^K = \frac{7}{12} \delta_K^2 $ &  $\tilde{\phi}_1^K = -\frac{7}{4} \delta_K^2 a_1^K$ & $\psi_1^K = \frac{7}{18} \delta_K^2$ & 
$\theta_1^K = \frac{7}{10} \delta_K^2 a_1^K$ & 
\\
$\phi_2^K = -\frac{7}{20} \delta_K^2 a_1^K$ & $\tilde{\phi}_2^K = \frac{7}{12} \delta_K^2$ & $\psi_2^K = \frac{7}{9} \delta_K^2$ & 
$\theta_2^K = -\frac{7}{5} \delta_K^2 a_1^K$ & 
\\
\hline
\end{tabular}
\\
\end{center}
\caption{\it Three-particle twist-4 parameter relations for kaon DA's derived from the renormalon model \protect{\cite{BBL,BGG}}.}
\label{tab-3}
\end{table}

\section{Parameters used in the calculation}

In this appendix we summarize the parameters used in the calculation of $f_{B_{(s)}K}$ form factors as well as in the calculation of 
$f_{B_{(s)}}$, Tables \ref{tab-1} - \ref{tab-3}. 
For DAs, the parameters and their $\mu$ dependence are taken from \cite{BBL,DKMMO}. 
The $\overline{MS}$ mass $\overline{m}_b$ entering the calculation is \cite{KSS}
\ba
\overline{m}_b( \overline{m}_b)= 4.164 \pm 0.025 \; \mbox{GeV}\,.
\label{eq:bmass}
\ea
The $\alpha_s(m_Z)$ value is the Particle Data Group (PDG) average \cite{PDG}. There is a new value for the $f_K$  decay constant \cite{RS}, 
prepared for the PDG's 2008 edition, 
\ba
f_{K} = (156 \pm 0.2 \pm 0.8 \pm 0.2)\, {\rm MeV}\, , 
\ea
which central value we adopt here, and the ratio of $K^-$ and $\pi^-$ decay constants is given by 
\ba
\frac{f_K}{f_{\pi}} = 1.196 \pm 0.002 \pm 0.006 \pm 0.001 \,.
\label{eq:ratio}
\ea
The $\overline{m}_s$ mass is the average \cite{CKP} of the QCD sum rule determinations from 
\cite{CK,JOP} and covers the $\overline{m}_s$ mass range given the most recently in \cite{DNRS}.
For the twist-2 kaon DA we use the first Gegenbauer moment $a_1^K$  calculated at NNLO accuracy from \cite{CKP}. 
Since the existing fits and calculations for the value of the second Gegenbauer moment $a_2^K$ show small SU(3) violation, 
with the large error, we accept here that $a_2^K = a_2^{\pi}$, with the value for $a_2^K$ given in Table 1 \cite{BBL}. 
We also use $m_K = 497.648 \pm 0.022$ MeV, $m_B =  5279.5 \pm 0.5$ MeV, and $m_{B_s} =  5366.1 \pm  0.6$ MeV \cite{PDG}. 

\end{document}